# Role of the Extra-Fe in $K_{2-x}Fe_{4+y}Se_5$ superconductors

Chih-Han Wang[1,2], Chih-Chien Lee[1], Gwo-Tzong Huang[2], Jie-Yu Yang[2], Ming-Jye Wang[3], Hwo-Shuenn Sheu[4], Jey-Jau Lee[4] and Maw-Kuen Wu[2]*

[1]Department of Electronic Engineering, National Taiwan University of Science and Technology, Taipei 10607, Taiwan
[2]Institute of Physics, Academia Sinica, Taipei 11529, Taiwan
[3]Institute of Astrophysics and Astronomy, Academia Sinica, Taipei 10617, Taiwan
[4]National Synchrotron Radiation Research Center, No. 101, Hsin-An Road, Hsinchu 30076, Taiwan

* E-mail: mkwu@phys.sinica.edu.tw

**Abstract**

The exact superconducting phase of $K_{2-x}Fe_{4+y}Se_5$ has yet conclusively decided since its discovery due to its intrinsic multiphase in early material. In an attempt to resolve the mystery, we have carried out systematic structural studies on a set of well controlled samples with exact chemical stoichiometry $K_{2-x}Fe_{4+x}Se_5$ (x=0–0.3) that are heat-treated at different temperatures. Our investigations, besides the determination of superconducting transition, focus on the detailed temperature evolution of the crystalline phases using high resolution synchrotron radiation X-ray diffraction. Our results show that superconductivity appears only in those samples been treated at high enough temperature and then quenched to room temperature. The volume fraction of superconducting transition strongly depends on the annealing temperature used. The most striking result is the observation of a clear contrast in crystalline phase between the non-superconducting parent compound $K_2Fe_4Se_5$ and the superconducting $K_{2-x}Fe_{4+y}Se_5$ samples. The x-ray diffraction patterned can be well indexed to the phase with I4/m symmetry in all temperature investigated. However, we need two phases with similar I4/m symmetry but different parameters to best fit the data at temperature below the Fe-vacancy order temperature. The results strongly suggest that superconductivity in $K_{2-x}Fe_{4+y}Se_5$ critically depends on the occupation of Fe atoms on the originally empty 4d site.

**Introduction**

The cuprates and Fe-based high Tc superconductors turn superconducting when their parent compounds are properly doped. The similarity of these superconductors, in which superconductivity emerges with the suppression of competing phases, has created the possibility for a unified picture of high temperature superconductivity [1, 2] The iron-based superconductors all have FeAs or FeSe layers, just as the $CuO_2$ layers, are the crucial ingredients for superconductivity [3-5]. It is generally believed the quasi-two-dimensional characteristics of these active layers and the proximity to magnetically ordered states induce superconductivity via unconventional pairing in these high Tc materials. However, the parent compounds of the Fe-pnictide superconductors are metallic with SDW antiferromagnetic order and with complex underlying electronic structure. This fact, in contrast to the Mott insulators in the cuprates, establishes an important difference between cuprates and pnictides. [6, 7].

On the other hand, the earlier works on FeSe superconductor considered this material to be Se-deficiency [8] so that the exact superconducting stoichiometry was $Fe_{1.01}Se$ [9]. Superconductivity in FeSe was found to closely associate with a structural transition [8, 10,11], which accompanies with a strongly enhanced spin fluctuation near Tc [12]. More spectroscopic studies [13-17] suggested modification of d-orbitals induced pseudogap might be the cause for the structural distortion, which is essential for superconductivity. Later studies indicate that there exists a parent compound [18], which in contrast to having excess Fe is actually Fe-deficient and is a Mott insulator. The material becomes superconducting after properly annealed this Fe-deficiency material at high temperature [19].

An important development in Fe-based superconductors is the discovery of the relatively higher Tc in alkaline-metal (A) intercalated FeSe with nominal composition $A_{0.8}Fe_2Se_2$ [20]. The nominal $A_{0.8}Fe_2Se_2$ sample was considered to crystallize in the tetragonal $ThCr_2Si_2$-type structure, which is isostructural to $BaFe_2As_2$ system. There were long-term debate regarding the exact superconducting phase in this material. One group claimed the superconducting composition is identified as $K_{0.83(2)}Fe_{1.64(1)}Se_2$ with enlarged √5x√5x1 crystallographic unit cell due to Fe-vacancy order [21]. The ideally Fe-vacancy order phase has a chemical stoichiometry of $K_{0.8}Fe_{1.6}Se_2$ ($K_2Fe_4Se_5$, referred to simply as ''245'' in the following), which order antiferromagnetically below 559K. The second group proposed that stoichiometric $K_xFe_2Se_2$ phase (which was indexed from x-ray diffraction to exist with about 13% in volume) is the superconducting phase in the matrix of the insulating AF 245 phase [22-25]. The third group suggested that superconductivity arises due to the interface between the iron vacancy ordered and free phase [26, 27]. Bao et al. has argued strongly that the antiferromagnetic order phase is the origin for superconducting phase [21,28], and slightly excess Fe is critical to the stability of the superconducting phase [29]. It was noted that the Fe-content of the superconducting samples $K_xFe_{1.6+y}Se_2$ is slightly off-stoichiometry compare from the parent compound 245 [30-34], which seems to play a significant character for the onset of superconductivity.

In our earlier study [35], we identified the ideally Fe-vacancy order 245 phase to be a Mott insulator with a block checkerboard antiferromagnetic (AFM) ordered below 559 K [21,36]. We also reported unambiguously that superconductivity could be induced with adding small amount of Fe to fill the vacant site plus strong disorder of the Fe-vacancy. It was noted that processing temperature has decisive effects on the properties of the samples. Superconducting transition only appears in those samples that have been annealed at high temperature and then quenched. Annealed the superconducting sample at low temperature (< 400C) gradually suppressed superconductivity with annealing time. The results showed that superconducting to non-superconducting state can switch back and forth in the same sample by changing the heat treatment process. We concluded that the observed facts strongly indicate superconductivity emerges as Fe-vacancy becomes disordered.

In order to understand better the effect of annealing conditions to superconductivity, we carried out detailed systematic study on a series of samples with stoichiometry $K_{2-x}Fe_{4+y}Se_5$ which subject to various annealing temperatures. We also took the advantage of the high-resolution x-ray diffraction in the National Synchrotron Radiation Research Center (NSRRC) with in-situ high temperature measurements. Our results revealed that the superconducting volume fraction, even with the same chemical stoichiometry, depends strongly on the annealing (and quench) temperature. There is a critical anneal temperature below which the superconducting volume fraction is close to zero. The high-temperature in-situ X-ray diffraction results further provide evidences that the random occupation of the original vacant-site is the key to superconductivity.

**Experimental**

Polycrystalline bulk samples were synthesized using a novel approach as described elsewhere in detail [35]. DC magnetic susceptibility measurements were performed in a Quantum Design superconducting quantum interference device vibrating sample magnetometer (SQUID-VSM). The Synchrotron X-ray diffraction of samples were performed at BL01C2, BL17A and TPS09A beamline of the NSRRC. The ring of NSRRC was operated at energy 1.5 GeV with a typical current 360 mA. The wavelength of the incident X-rays was 0.61993Å, 1.3216Å and 0.77491Å, delivered from the 5 Tesla Superconducting Wavelength Shifter and a Si(111) triangular crystal monochromator. Two pairs of slits and one collimator were set up inside the experimental hutch to provide a collimated beam with dimensions of typical 0.5 mm x 0.5 mm (H x V) at the sample position. The diffraction pattern was recorded with a Mar345 imaging plate detector approximately 400 mm from the sample and typical exposure duration 1 min. The pixel size of Mar345 was 100 m. The one-dimensional powder diffraction profile was converted with program GSAS II and cake-type integration. The diffraction angles were calibrated according to Bragg positions of LaB6 standards.

**Results and Discussion**

Figure 1 (a) shows the X-ray diffraction pattern of three $K_{2-x}Fe_{4+x}Se_5$ samples with x ranging from 0.1 to 0.2. The inset displays the relevant magnetic susceptibility of the samples. The magnetic data indicates the superconducting diamagnetic signal diminishes with reducing K-content, almost no diamagnetic signal in $K_{1.8}Fe_{4.2}Se_5$. Meanwhile the X-ray diffraction patterns show no additional peaks, which have been associated with the extra $K_zFe_2Se_2$ phase by other groups [22], in this non-superconducting $K_{1.8}Fe_{4.2}Se_5$ sample. Based on this observation one would consider superconductivity might originate from this extra $K_zFe_2Se_2$ phase. However, examining closely the details X-ray and magnetic data of two $K_{1.9}Fe_{4.2}Se_5$ samples, as shown in figure 1 (b), which are prepared under different annealing and quenching temperatures, one at 650°C, the other at 790°C, it shows that high volume superconducting signal in the sample prepared at 790°C, but no diamagnetic signal in sample prepared at 650°C. The X-ray diffraction patterns for both samples are almost identical showing the presence of the extra peaks associated with $K_zFe_2Se_2$ phase. The above contradictory observations suggest that there are more subtleties regarding the correlation between crystal structure and superconductivity in this intriguing superconducting system.

Subsequently, we carried out detailed X-ray diffractions at different temperatures up to 750°C for the $K_2Fe_4Se_5$ (2(4)5) and excess-Fe $K_{1.9}Fe_{4.2}Se_5$ (2(4.2)5) samples. Both the warming and cooling cycles exhibit no difference in the diffraction patterns, as shown in Figure 2. Figure 2(a) illustrates the significant shifts in diffraction peaks (008) and (051) at 275°C in warming, which are the indication of the transition from the Fe-vacancy order to disorder state (see fig. 5(a)). A large hysteresis showing the transition temperature was reduced to 240°C in cooling cycle. Below the transition temperature, the material exhibits clearly Fe-vacancy order of I4/m symmetry with $\sqrt{5} \times \sqrt{5}$ superlattice; and above the order temperature the crystal symmetry seems to fit well to I4/mmm symmetry. In addition, the super-structural peaks related to the vacancy order state completely disappear above the ordering temperature. However, since the chemical stoichiometry of the sample is exactly 245, it was a puzzle to us how can the high temperature phase belong to the high symmetry I4/mmm space group as there is 20% Fe-deficiency? Thus, we tried to fit the diffraction pattern with I4/m symmetry considering that the high temperature phase Fe could occupy both 16*i* (fully occupied) and 4*d* site (originally empty site). The fitting was almost perfect and the resulted lattice parameters are shown in supplementary result table SI, which is almost identical to that fitted with I4/mmm. This result indicates the crystal structure of 245 could be well described by single phase with I4/m symmetry even above 275°C [29].

It was to our surprise when we first saw the temperature dependent diffraction patterns of the excess-Fe sample that exhibits bulk superconductivity, as shown in figure 2(b), which show essentially no shift in all diffraction peak positions (main phase) throughout the whole temperature range. More carefully examined the diffraction patterns, as shown in figure (2b) and 3, some additional peaks appear below the vacancy-order temperature at 270°C during warming. The data for 2(4.2)5 sample are found to best fit with two different sets of parameters, one with Fe-vacancy ordered and the other disordered Fe-vacancy, under the same I4/m symmetry (as shown in supplementary result table SII). Figure 2 (c) and 2 (e) plot the lattice parameters of 245 and 2(4.2)5 samples, respectively, below and above the Fe-vacancy order temperature (both fit with I4/m symmetry). Figure 2(d) and (f) are the schematic atomic arrangements for the two samples based on the refined results.

Figure 3 (a) and (b) display the diffraction patterns for 245 and 2(4.2)5 samples plotting with diffraction angles and d-spacing, respectively. There are two phases coexisting in excess-Fe sample shown in figure 3 (a). Inset shows the superlattice (110) profile which revealing two phases in1.9(4.2)5. one with a longer in-plane lattice parameters and the other with a smaller lattice parameter with I4/m symmetry. From figure 3(b), those additional features can be associated with the Miller index (002), (130), (132), (134), (136) … Fe-layers, based on I4/m symmetry. Figure 3(c) show the schematic atomic arrangement of the plane (132) in the vacancy ordered and disordered states. The d-spacing for disordered state is about 2% larger than that of the ordered state. These planes are all crossing the Fe-4*d* site, as exemplified by the schematic plot in figure 3 (d). This result indicates that the presence of excess-Fe atoms shall begin to fill the original empty 4*d* sites (construct the disorder phase). Thus, additional planes with the same Miller index show up. Our analysis show that the difference in d-spacing of these new features, displayed in figure 3 (b), with the original peaks is at most 2%. These additional features disappear above the vacancy order-disorder transition temperature, but keep the diffraction peak positions (Fe-vacancy disordered phase) unchanged (see fig. 2(b)). This observation suggests that the added excess-Fe atoms play critical role to maintain the crystal lattice (size) with the I4/m symmetry as that of the vacancy ordered state. Above the vacancy order temperature, the lattice remains with the I4/m symmetry but with disorder occupation of the Fe-atom to all possible Fe-site so that the additional features (Fe-vacancy ordered phase) disappear.

We have also tried to fit the observed diffraction patterns by considering whether there exists an orthorhombic symmetry as it is known that for tetragonal symmetry, the peak of (hkl) and (khl) are at same diffraction angle because $a=b$, whereas for orthorhombic symmetry, the peak of (hkl) and (hkl) could be at different diffraction angle because $a \neq b$. Detailed analysis of the data suggests that a monoclinic 245 structure with I112/m symmetry, using the lattice parameters: $a$ = 8.6Å, b = 8.72Å, c = 14.2Å and γ = 90°, could generate most of the diffraction peaks observed (as shown in supplementary result figure S1). Though the diffraction peaks seem to fit well with the observed results, the overall fit was relatively poor regarding the peak intensity, which is known to depend on the number, position, and specie of atoms.

Based on the above observations, one might expect even in the stoichiometry 245 sample to observe Fe atoms occupied the empty 4*d* site if the sample has been treated at high enough temperature so that Fe atoms could hop around. Indeed, the high-resolution x-ray diffraction patterns of the 245 sample quenched from 800°C, as shown in figure 4, display extra peaks similar to those observed in excess-Fe samples (refined lattice parameters are shown in supplementary result table SIII). Meanwhile, the magnetic susceptibility data shown in the inset of figure 4 (reproduced with permission from ref. 35) exhibits clearly diamagnetic signal indicating the presence of superconductivity though its volume fraction is rather small. These results demonstrate that the occupation of the 4*d*-site meanwhile maintain the crystal lattice (size) plays a key role for the emergence of superconductivity. It is noted that there is a significant temperature hysteresis in the vacancy order-disorder transition as on cooling the additional features do not appear until below 240°C suggesting this transition is first order thermodynamically.

To clarify further the role of the extra Fe-atom to the crystal lattice, two more samples were prepared for detailed studies: $K_{1.9}Fe_{4.05}Se_5$ and $K_{1.9}Fe_{4.25}Se_5$. As shown in Fig. 5, the changes in diffraction peak positions of (0 0 8) and (0 5 1) for $K_{1.9}Fe_{4.05}Se_5$ sample clearly are less significant comparing with that of the 245 sample; and for $K_{1.9}Fe_{4.25}Se_5$ sample, essentially the same as that of the $K_{1.9}Fe_{4.2}Se_5$ sample, the peak positions do not change with temperature. These results further confirm that extra-Fe atoms begin to fill to the empty 4*d*-site and stabilize the structural phase I4/m symmetry so that the lattice size maintain the same above the order-disorder temperature. By comparing with the parent compound, the extra-Fe sample exhibits much weaker super-structural peaks such as the (222) peak, which almost disappears at room temperature. This demonstrated the addition of extra-Fe suppresses the Fe-vacancy long range order. Consequently, this modification suppresses the magnetic order, which accompanies with the vacancy order, to favor the emergence of superconductivity.

Our experimental observations further confirm that superconductivity in $K_{2-x}Fe_{4+y}Se_5$ is not due to the impurity phase ($K_{0.5}Fe_2Se_2$), which have been suggested based on the x-ray diffraction features that coincide with what we observed in the excess-Fe samples. Our results unambiguously demonstrated that the random occupation of Fe-atom in the lattice is key for superconductivity. Nevertheless, the exact origin for superconductivity remains to be resolved. Several questions remain to be answered, such as why superconducting volume fraction in the excess-Fe sample depends on the annealing and quenching temperature? We are currently working on the detailed refinement of the X-ray diffraction data that collected with wide temperature range. A very preliminary result indicates that a more subtle high temperature structural distortion might be closely related to the appearance of superconductivity. Details of these results will be published in the near future.

## Summary

In summary, in this detailed X-ray diffraction experiment over wide-range of temperature, we have showed that the superconducting $K_{2-x}Fe_{4+y}Se_5$ samples with excess-Fe atoms exhibit the same structure with I4/m symmetry throughout the whole temperature range studied. The observation of extra features, which have been considered by others as the presence of impurity phase associated with the I4/mmm symmetry, actually were an in-plane compressed I4/m lattice with Fe-vacancy ordering. The occupation of the Fe-4$d$ site and eventually more randomly distributed of the Fe-atom are critical to the emergence of superconductivity. Therefore, the main conclusion of this study further confirms that superconductivity in $K_{2-x}Fe_{4+y}Se_5$, very much similar to that observed in high Tc cuprates, derived from the doping (add extra Fe) to the parent Mott insulator $K_2Fe_4Se_5$. The more disorder of the Fe occupation in the I4/m lattice leads to higher volume fraction of superconductivity. A more detailed refinement of the high temperature diffraction data is currently undergoing extensively in order to unravel the exact origin for the appearance of superconductivity in this intriguing FeSe based superconductor.

## Acknowledgments

We thank Chung-Kai Chang for fruitful discussion and valuable suggestions in Synchrotron PXRD measurement. We appreciate the help very much from Dr. Phillip Wu in preparing this manuscript. This research is supported by the Ministry of Science and Technology Grant MOST106-2633M-001-001, Academia Sinica Thematic Research, Taiwan and National Synchrotron Radiation Research Center, Taiwan

## References:

1. David C. Johnston, Adv. Phys. 59, 803 (2010).
2. Johnpierre Paglione and Richard L. Greene, Nat. Phys. 6, 645 (2010).
3. Elbio Dagotto, Rev. Mod. Phys. 66, 763 (1994).
4. Scalapino, D. J., Phys. Rep. 250, 329 (1995).
5. Elbio Dagotto, Rev. Mod. Phys. 85, 849 (2013).
6. P J Hirschfeld, M M Korshunov, and I I Mazin, Rep. Prog. Phys. 74, 124508 (2011).
7. G. R. Stewart, Rev. Mod. Phys. 83, 1589 (2011).
8. Fong-Chi Hsu, Jiu-Yong Luo, Kuo-Wei Yeh, Ta-Kun Chen, Tzu-Wen Huang, Phillip M. Wu, Yong-Chi Lee, Yi-Lin Huang, Yan-Yi Chu, Der-Chung Yan, and Maw-Kuen Wu, Superconductivity in the PbO-type structure α-FeSe, Proc. Natl. Acad. Sci. U.S.A. 105, 14262 (2008).
9. T. M. McQueen, Q. Huang, V. Ksenofontov, C. Felser, Q. Xu, H. Zandbergen, Y. S. Hor, J. Allred, A. J. Williams, D. Qu, J. Checkelsky, N. P. Ong, and R. J. Cava, Extreme sensitivity of superconductivity to stoichiometry in $Fe_{1+\delta}Se$, Phys. Rev. B 79, 014522 (2009).
10. M. J. Wang, J.Y. Luo, T.W. Huang, H. H. Chang, T. K. Chen, F. C. Hsu, C. T. Wu, P. M. Wu, A. M. Chang, and M. K. Wu, Crystal Orientation and Thickness Dependence of the Superconducting Transition Temperature of Tetragonal $FeSe_{1-x}$ Thin Films, Phys. Rev. Lett. 103, 117002 (2009).
11. T. M. McQueen, A. J.Williams, P.W. Stephens, J. Tao, Y. Zhu, V. Ksenofontov, F. Casper, C. Felser, and R. J. Cava, Tetragonal-to-Orthorhombic Structural Phase Transition at 90K in the Superconductor $Fe_{1.01}Se$, Phys. Rev. Lett. 103, 057002 (2009).
12. T. Imai, K. Ahilan, F. L. Ning, T. M. McQueen, and R. J. Cava, Why Does Undoped FeSe Become a High-Tc Superconductor under Pressure? Phys. Rev. Lett. **102**, 177005 (2009)
13. Y.-C. Wen, K.-J. Wang, H.-H. Chang, J.-Y. Luo, C.-C. Shen, H.-L. Liu, C.-K. Sun, M.-J. Wang, and M.-

K. Wu, Gap Opening and Orbital Modification of Superconducting FeSe above the Structural Distortion, Phys. Rev. Lett. 108, 267002 (2012).
14. A. M. Zhang, J. H. Xiao, Y. S. Li, J. B. He, D. M. Wang, G. F. Chen, B. Normand, and Q. M. Zhang, Two-magnon Raman scattering in $A_{0.8}Fe_{1.6}Se_2$ systems (A = K, Rb, Cs, and Tl): Competition between superconductivity and antiferromagnetic order, Phys. Rev. B 85, 214508 (2012).
15. A. M. Zhang, K. Liu, J. B. He, D. M. Wang, G. F. Chen, B. Normand, and Q. M. Zhang, Effect of iron content and potassium substitution in A0.8Fe1.6Se2 (A=K, Rb, Tl) superconductors: A Raman scattering investigation, Phys. Rev. B 86, 134502 (2012).
16. M. Yi, D. H. Lu, R. Yu, S. C. Riggs, J.-H. Chu, B. Lv, Z. K. Liu, M. Lu, Y.-T. Cui, M. Hashimoto, S.-K. Mo, Z. Hussain, C.W. Chu, I. R. Fisher, Q. Si, and Z.-X. Shen, Observation of Temperature-Induced Crossover to an Orbital-Selective Mott Phase in $A_xFe_{2-y}Se_2$ (A=K, Rb) Superconductors, Phys. Rev. Lett. 110, 067003 (2013).
17. J. Maletz, V. B. Zabolotnyy, D. V. Evtushinsky, S. Thirupathaiah, A. U. B. Wolter, L. Harnagea, A. N.Yaresko, A. N. Vasiliev, D. A. Chareev, A. E. B¨ohmer, F. Hardy, T. Wolf, C. Meingast, E. D. L. Rienks, B. B¨uchner, and S. V. Borisenko, Unusual band renormalization in the simplest iron-based superconductor $FeSe_{1-x}$, Phys. Rev. B 89, 220506(R) (2014).
18. Ta-Kun Chena, Chung-Chieh Chang, Hsian-Hong Chang, Ai-Hua Fang, Chih-Han Wang, Wei-Hsiang Chao, Chuan-Ming Tseng, Yung-Chi Lee, Yu-Ruei Wu, Min-Hsueh Wen, Hsin-Yu Tang, Fu-Rong Chen, Ming-Jye Wang, Maw-Kuen Wu, and Dirk Van Dyck, Fe-vacancy order and superconductivity in tetragonal β-$Fe_{1-x}Se$, Proc. Natl. Acad. Sci. U.S.A. 111, 63 (2014).
19. Chung-Chieh Chang, Chih-Han Wang, Min-Hsueh Wen, Yu-Ruei Wu, Yao-Tsung Hsieh, Maw-Kuen Wu, Superconductivity in PbO-type tetragonal FeSe nanoparticles, Solid State Communications 152, 649 (2012).
20. Jiangang Guo, Shifeng Jin, Gang Wang, Shunchong Wang, Kaixing Zhu, Tingting Zhou, Meng He, and Xiaolong Chen, Superconductivity in the iron selenide $K_xFe_2Se_2$ (0≦ x≦1.0), Phys. Rev. B 82, 180520(R) (2010).
21. BAO Wei, HUANG Qing-Zhen, CHEN Gen-Fu, M. A. Green, WANG Du-Ming, HE Jun-Bao, QIU Yi-Ming, A Novel Large Moment Antiferromagnetic Order in K0.8Fe1.6Se2 Superconductor, CHIN. PHYS. LETT. 28, 086104 (2011).
22. Scott V. Carr, Despina Louca, Joan Siewenie, Q. Huang, Aifeng Wang, Xianhui Chen, and Pengcheng Dai, Structure and composition of the superconducting phase in alkali iron selenide $K_yFe_{1.6+x}Se_2$, Phys. Rev. B 89, 134509 (2014).
23. Daniel P. Shoemaker, Duck Young Chung, Helmut Claus, Melanie C. Francisco, Sevda Avci, Anna Llobet, and Mercouri G. Kanatzidis, Phase relations in $K_xFe_{2-y}Se_2$ and the structure of superconducting $K_xFe_2Se_2$ via high-resolution synchrotron diffraction, Phys. Rev. B 86, 184511 (2012).
24. Z. Wang, Y. Cai, Z. W. Wang, C. Ma, Z. Chen, H. X. Yang, H. F. Tian, and J. Q. Li, Archimedean solidlike superconducting framework in phase-separated $K_{0.8}Fe_{1.6+x}Se_2$ (0≦x≦0.15), Phys. Rev. B 91, 064513 (2015).
25. A. Ricci, N. Poccia, B. Joseph, D. Innocenti, G. Campi, A. Zozulya, F. Westermeier, A. Schavkan, F. Coneri, A. Bianconi, H. Takeya, Y. Mizuguchi, Y. Takano, T. Mizokawa, M. Sprung, and N. L. Saini, Phys. Rev. B 91, 020503 (2015).
26. Wei Li, Hao Ding, Zhi Li, Peng Deng, Kai Chang, Ke He, Shuaihua Ji, Lili Wang, Xucun Ma, Jiang-Ping Hu, Xi Chen, and Qi-Kun Xue, $KFe_2Se_2$ is the Parent Compound of K-Doped Iron Selenide Superconductors, Phys. Rev. Lett. 109, 057003 (2012).
27. Chunruo Duan, Junjie Yang, Yang Ren, and Despina Louca, Superconductivity at the vacancy disorder boundary in $K_xFe_{2-y}Se_2$, arXiv:1706.02182v2.
28. F. Ye, S. Chi, Wei Bao, X. F. Wang, J. J. Ying, X. H. Chen, H. D. Wang, C. H. Dong, and Minghu Fang, Common Crystalline and Magnetic Structure of Superconducting $A_2Fe_4Se_5$ (A = K,Rb,Cs,Tl) Single Crystals Measured Using Neutron Diffraction, Phys. Rev. Lett. 107, 137003 (2011).
29. Wei Bao, Structure, magnetic order and excitations in the 245 family of Fe-based superconductors, J. Phys.: Condens. Matter 27, 023201 (2015).
30. Zhi-Wei Wang, Zhen Wang, Yuan-Jun Song, Chao Ma, Yao Cai, Zhen Chen, Huan-Fang Tian, Huai-Xin Yang, Gen-Fu Chen, and Jian-Qi Li, Structural Phase Separation in $K_{0.8}Fe_{1.6+x}Se_2$ Superconductors, J. Phys. Chem. C 116, 17847 (2012).
31. F. Peng, W.P. Liu, C.T. Lin, Study of Thermal Behavior and Single Crystal Growth of $A_{0.8}Fe_{1.81}Se_2$ (A = K, Rb, and Cs), J Supercond Nov Magn 26, 1205 (2013).

32. Yong Liu, Qingfeng Xing, Warren E. Straszheim, Jeff Marshman, Pal Pedersen, Richard McLaughlin, and Thomas A. Lograsso, Formation mechanism of superconducting phase and its three-dimensional architecture in pseudo-single-crystal $K_xFe_{2-y}Se_2$, Phys. Rev. B 93, 064509 (2016).
33. J. Yang, C. Duan, Q. Huang, C. Brown, J. Neuefeind, and Despina Louca, Strong correlations between vacancy and magnetic ordering in superconducting $K_{0.8}Fe_{2-y}Se_2$, Phys. Rev. B 94, 024503 (2016).
34. A. Ricci, N. Poccia, G. Campi, B. Joseph, G. Arrighetti, L. Barba, M. Reynolds, M. Burghammer, H. Takeya, Y. Mizuguchi, Y. Takano, M. Colapietro, N. L. Saini, and A. Bianconi, Nanoscale phase separation in the iron chalcogenide superconductor $K_{0.8}Fe_{1.6}Se_2$ as seen via scanning nanofocused x-ray diffraction, Phys. Rev. B 84, 060511(R) (2011).
35. Chih-Han Wang, Ta-Kun Chen, Chung-Chieh Chang, Chia-Hao Hsu, Yung-Chi Lee, Ming-Jye Wang, Phillip M. Wu, and Maw-Kuen Wu, Disordered Fe vacancies and superconductivity in potassium-intercalated iron selenide ($K_{2-x}Fe_{4+y}Se_5$), Europhys. Lett. 111, 27004 (2015).
36. Bao Wei, Physics picture from neutron scattering study on Fe-base superconductors, Chin. Phys. B 22, 087405 (2013).

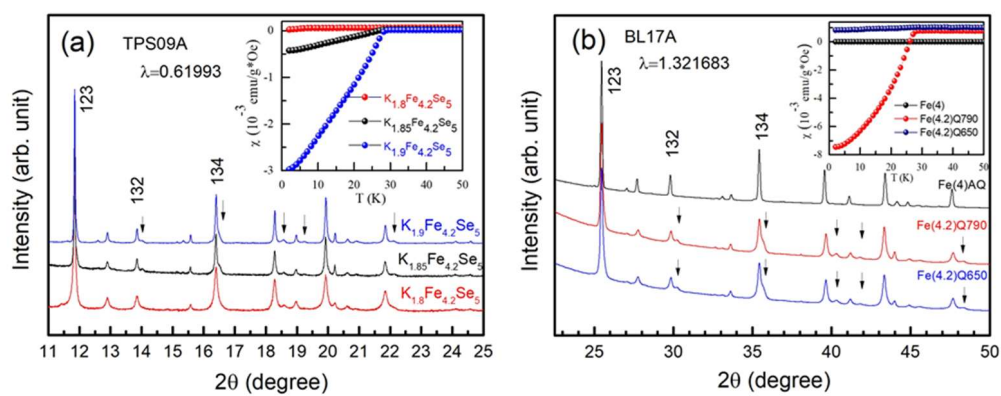

**Fig. 1. XRD patterns of polycrystalline $K_{2-x}Fe_{4+y}Se_5$ Samples.** (a) $K_{2-x}Fe_{4.2}Se_5$ quenched from 820°C. The marks point out the second phase. When K content decreases, the second phase gradually disappears. (b) $K_{1.9}Fe_{4.2}Se_5$ quenched in 650°C and 790°C. Same composition with same structure but shows no superconductivity quenched from 650°C. The inset is the magnetic susceptibility measured in a 30 Oe field under zero-field-cooled (ZFC) conditions.

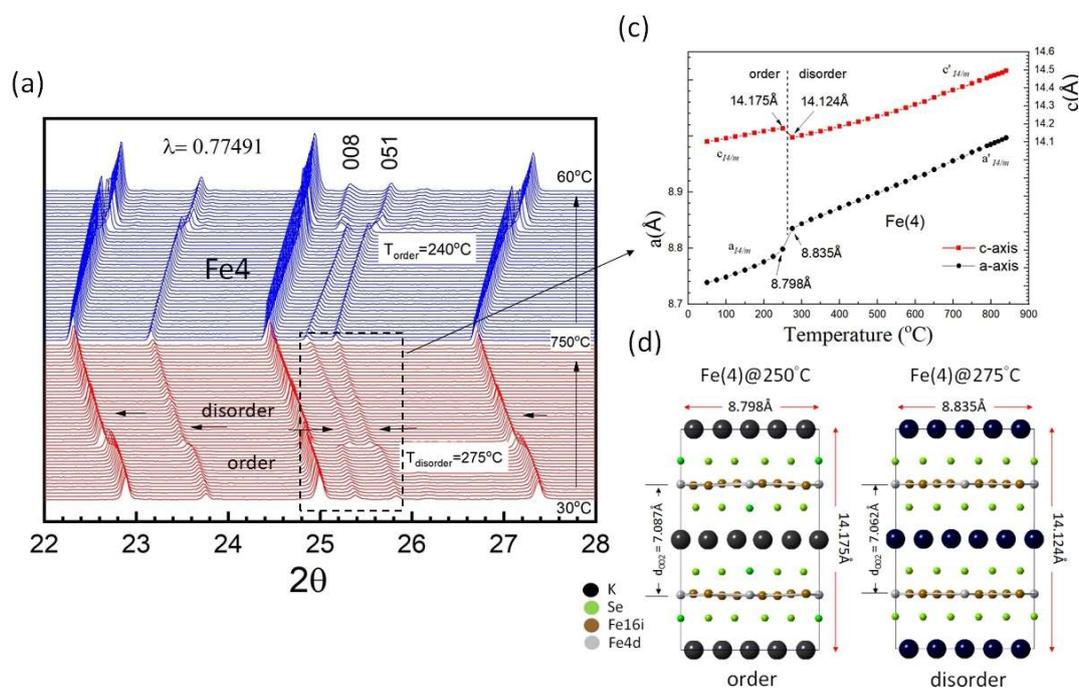

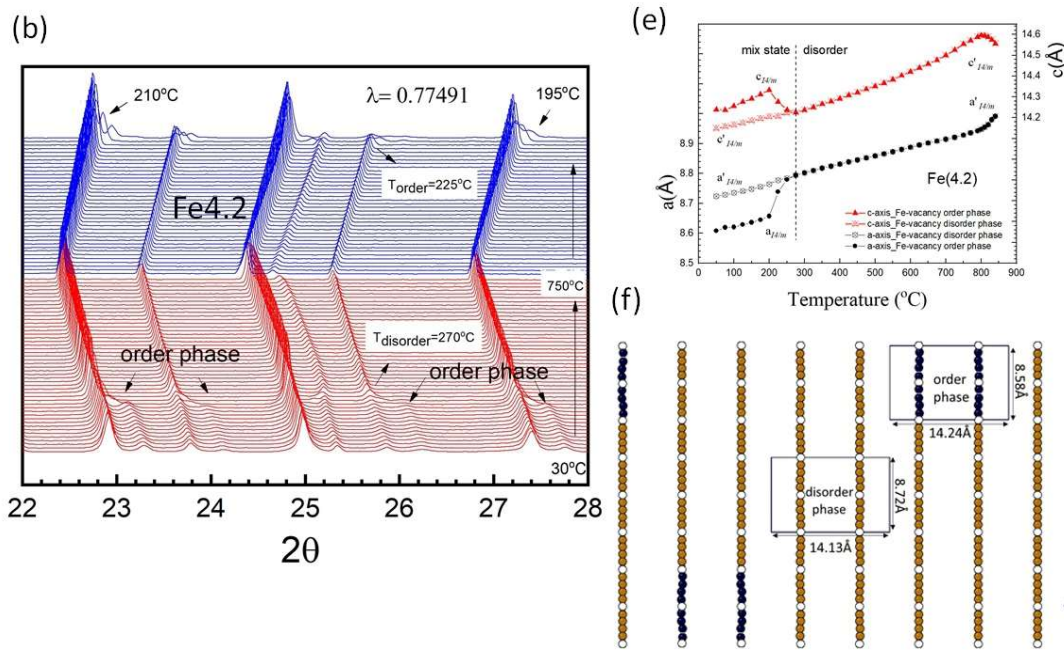

**Fig. 2. Evolution of the temperature dependence x-ray powder diffraction pattern of** (a) $K_2Fe_4Se_5$ (245) (b) $K_{1.9}Fe_{4.2}Se_5$ (2(4.2)5). Color red indicate increasing temperature; blue indicates decreasing temperature. Refined lattice parameters of $K_2Fe_4Se_5$ and $K_{1.9}Fe_{4.2}Se_5$ samples are displayed in figure (c) and (e) respectively. All data are refined using the I4/m unit cell. A cell with compressed c-axis and expanded ab-plane is observed above vacancy disorder temperature in 245 sample; two cells with different lattice parameters, one for vacancy-ordered cell the other is vacancy disordered, are found to best fit the data for 2(4.2)5 sample below Fe-vacancy disorder temperature. (d) Illustrated the I4/m unit cell. Left cell is Fe-vacancy ordered phase and right cell is disordered phases. In the Fe-vacancy order phase, the mirror image of Fe2 (16i) layers is according to magnetic moment orientation [21]. In disordered phase shows the c-axis is compressed, and ab-plane is expanded. (f) Schematic of mix structure which included both ordered and disordered phases. The black boxes outline the I4/m unit cell. The iron vacancy-site (4d) is marked by the hollow symbol and Fe(16i) site is marked by the khaki (black) color. K and Se atoms are not shown in the plot.

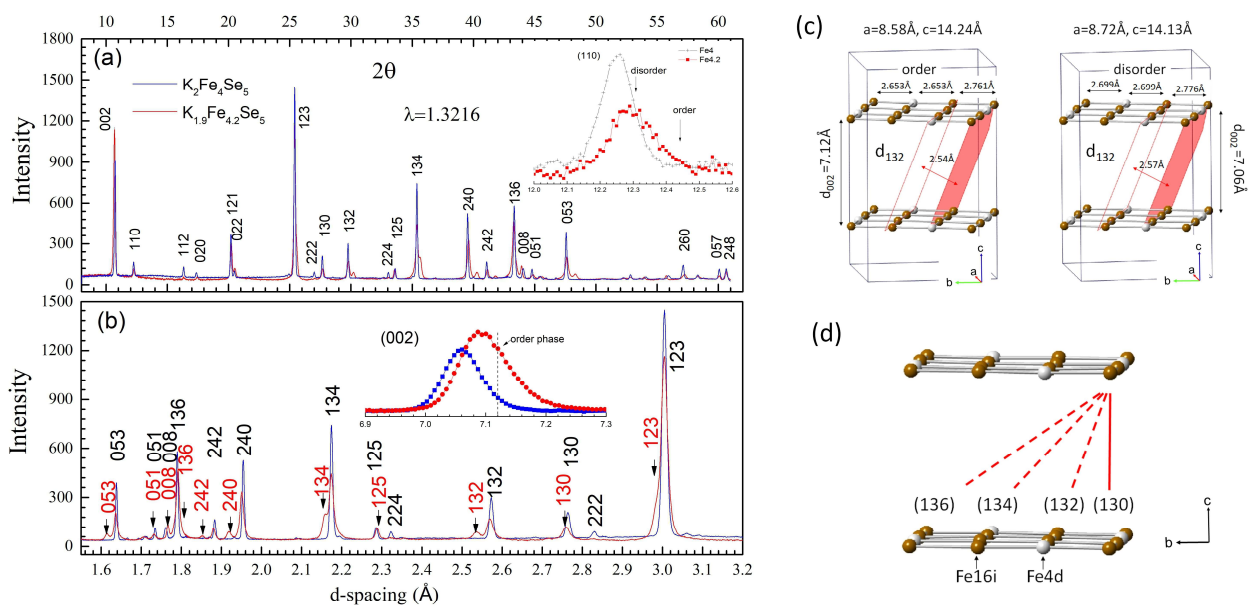

**Figure 3. Synchrotron powder diffraction patterns of the polycrystalline $K_{2-x}Fe_{4+y}Se_5$ Samples.** (a) Superconducting

1.9(4.2)5 and non-superconducting parent 245 x-ray diffraction pattern in diffraction angles. The superlattice peaks, e.g., (222) and (224) peaks, are almost suppressed in the superconducting 1.9(4.2)5 sample. Inset shows the (110) diffraction profile. (b) The same diffraction patterns displayed with lattice d-spacing show that there are two similar phases in 1.9(4.2)5 sample. One is similar to 245 parent phase but without (222) superlattice, and the other with slightly smaller d-spacing analyze. Inset shows the (002) profile. (c) illustrate the two similar phases with difference d-spacing for (132) plane. (d) Schematic of (130), (132), (134) and (136) plane. The iron vacancy site 4d is marked by white symbol and the 16i site is marked using khaki color.

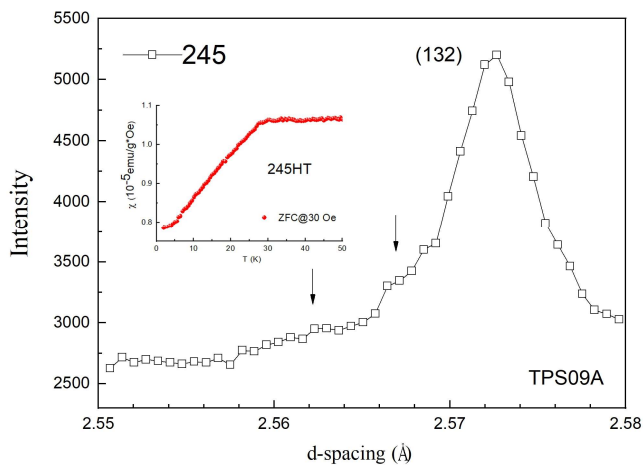

**Figure 4. Synchrotron powder diffraction patterns of the polycrystalline 245HT Sample.** Diffraction profile of the (132) plane is shown. The multi-phase profile of 245HT is similar to those observed in excess-Fe samples. Inset displays a clear diamagnetic signal at 29.6K.

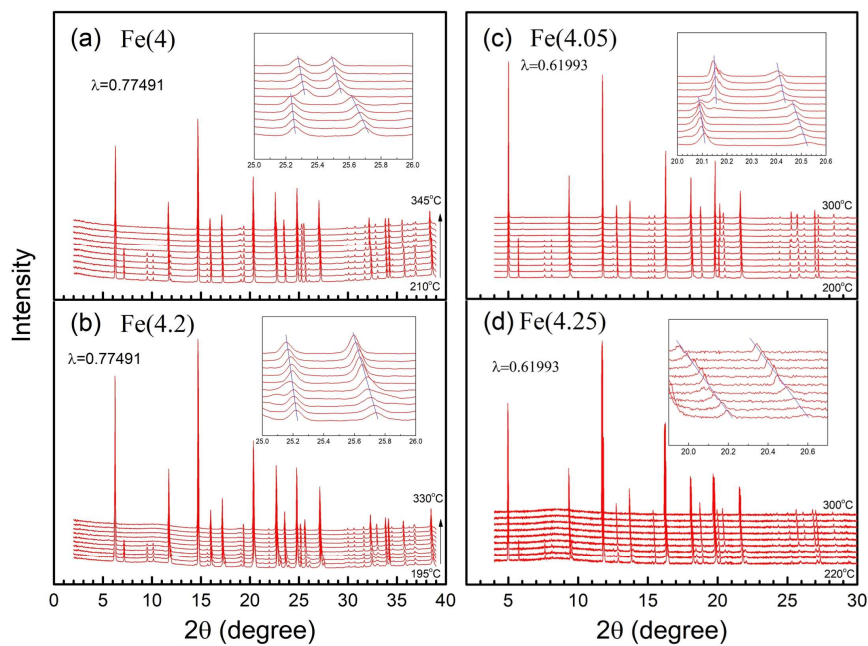

**Fig. 5. Temperature dependence in situ x-ray powder diffraction pattern of** (a) $K_2Fe_4Se_5$ (b) $K_{1.9}Fe_{4.2}Se_5$ (c) $K_{1.9}Fe_{4.05}Se_5$ (d) $K_{1.9}Fe_{4.25}Se_5$. The inset enlarges 2θ within the range 25° < 2θ < 26° and 20° < 2θ < 21° (depends on wavelength λ). The pictures show when temperature higher than Ts (a) Fe(4) lattice constant has obvious change (c) Fe(4.05) lattice constant has minor change (b) Fe(4.2) and (d) Fe(4.25) almost the same. This indicates excess Fe stabilizes the 245 structure.

Supplementary Materials

Table SI: The structural information for $K_2Fe_4Se_5$ as Rietveld refinement from the synchrotron diffraction data.

| Space group: I4/m | | | | | 2(Fe4)5-275°C | Space group: I4/mmm | | | | |
|---|---|---|---|---|---|---|---|---|---|---|
| a (Å) = 8.834865 ; c (Å) = 14.123927   $\alpha=\beta=\gamma=90°$ | | | | | | a (Å) = 3.951249 ; c (Å) = 14.124636   $\alpha=\beta=\gamma=90°$ | | | | |
| | x | y | z | mult. | Occ. | | x | y | z | mult. | Occ. |
| $K_1$(2b) | 0 | 0 | 0.5 | 2 | 0.9206 | $K_1$(2b) | 0 | 0 | 0 | 2 | 0.8044 |
| $K_2$(8h) | 0.797342 | 0.403553 | 0.5 | 8 | 0.7817 | $Fe_1$(4d) | 0 | 0.5 | 0.25 | 4 | 0.7980 |
| $Fe_1$(4d) | 0 | 0.5 | 0.25 | 4 | 0.7989 | $Se_1$(4e) | 0 | 0 | 0.354242 | 4 | 1.0017 |
| $Fe_2$(16i) | 0.300480 | 0.396745 | 0.249143 | 16 | 0.8022 | | | | | | |
| $Se_1$(4e) | 0 | 0 | 0.147326 | 4 | 1.0709 | | | | | | |
| $Se_2$(16i) | 0.100799 | 0.299760 | 0.354629 | 16 | 0.9896 | | | | | | |
| Rwp = 5.40% , Rp = 3.93% , $x^2$ = 0.09295 | | | | | | Rwp = 5.47% , Rp = 4.01% , $x^2$ = 0.09351 | | | | | |

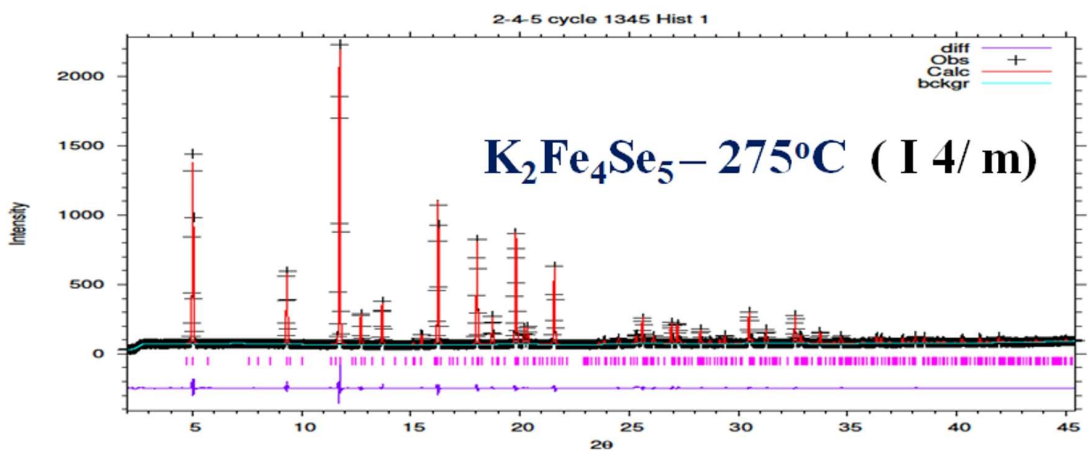

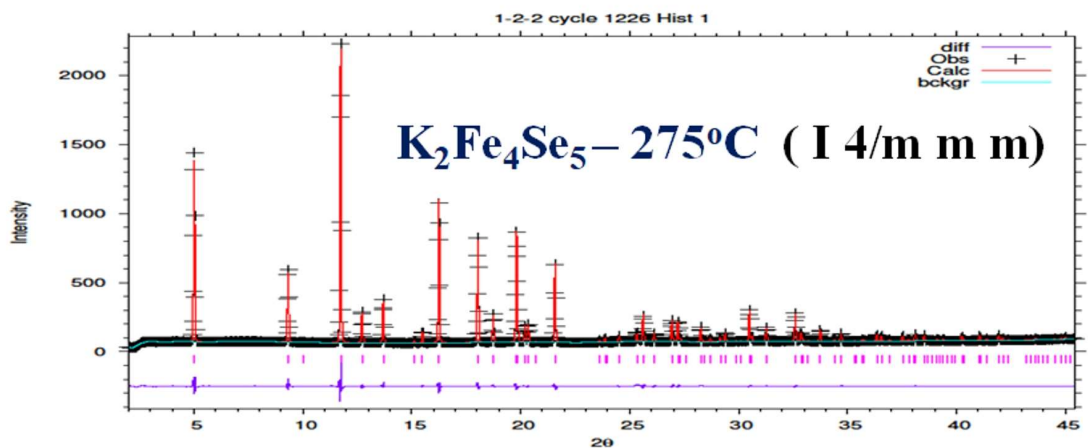

Table SII: The structural information for $K_{1.9}Fe_{4.2}Se_5$ as Rietveld refinement from the synchrotron diffraction data.

| Fe-vacancy disorder phase (86.3%) | | | | | 1.9(Fe4.2)5-25°C | Fe-vacancy order phase (13.7%) | | | | |
|---|---|---|---|---|---|---|---|---|---|---|
| a (Å) = 8.717743 ; c (Å) = 14.1251740 α=β=γ=90° | | | | | | a (Å) = 8.582171 ; c (Å) = 14.235626 α=β=γ=90° | | | | |
| | x | y | z | mult. | Occ. | | x | y | z | mult. | Occ. |
| $K_1$(2b) | 0 | 0 | 0.5 | 2 | 0.8509 | $K_1$(2b) | 0 | 0 | 0.5 | 2 | 0.7841 |
| $K_2$(8h) | 0.808077 | 0.392152 | 0.5 | 8 | 0.8074 | $K_2$(8h) | 0.732769 | 0.398093 | 0.5 | 8 | 0.6877 |
| $Fe_1$(4d) | 0 | 0.5 | 0.25 | 4 | 0.2214 | $Fe_1$(4d) | 0 | 0.5 | 0.25 | 4 | 0.0000 |
| $Fe_2$(16i) | 0.295179 | 0.406466 | 0.250104 | 16 | 1.0083 | $Fe_2$(16i) | 0.293777 | 0.403941 | 0.252191 | 16 | 1.0315 |
| $Se_1$(4e) | 0 | 0 | 0.137406 | 4 | 1.0006 | $Se_1$(4e) | 0 | 0 | 0.144785 | 4 | 1.0324 |
| $Se_2$(16i) | 0.110745 | 0.292429 | 0.353529 | 16 | 0.9878 | $Se_2$(16i) | 0.101194 | 0.302082 | 0.357016 | 16 | 0.9602 |
| Rwp = 2.94% ; Rp = 2.19% ; $\chi^2$ = 0.05283 | | | | | | Space group: I4/m | | | | |

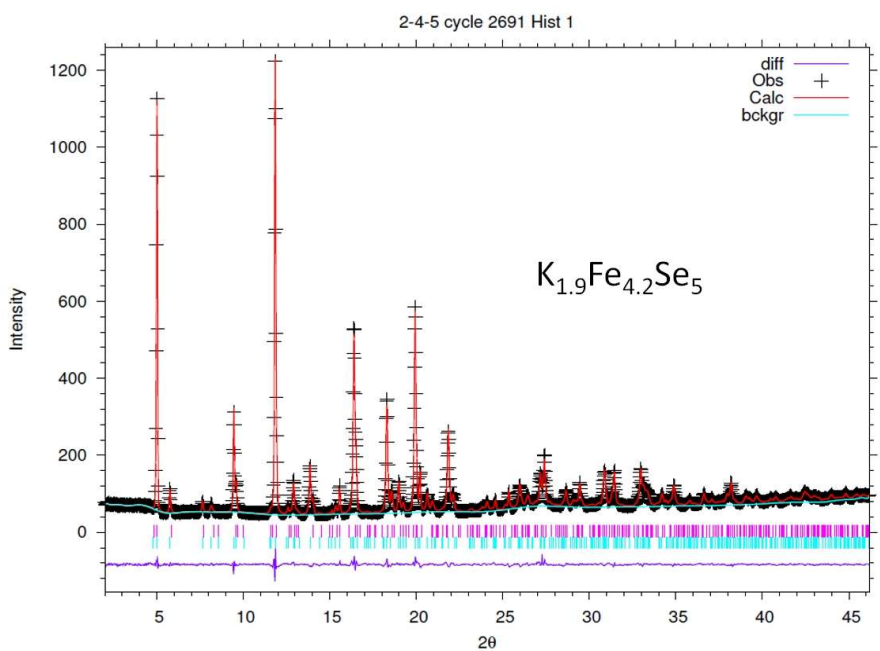

Table SIII: The structural information for 245HT as Rietveld refinement from the synchrotron diffraction data.

| Fe-vacancy disorder phase (16.83%) | | | | | 245HT | Fe-vacancy order phase (83.17%) | | | | |
|---|---|---|---|---|---|---|---|---|---|---|
| a (Å) = 8.831810 ; c (Å) = 13.974562 α=β=γ=90° | | | | | | a (Å) = 8.732868 ; c (Å) = 14.115056 α=β=γ=90° | | | | |
| | x | y | z | mult. | Occ. | | x | y | z | mult. | Occ. |
| $K_1$(2b) | 0 | 0 | 0.5 | 2 | 0.8298 | $K_1$(2b) | 0 | 0 | 0.5 | 2 | 0.8828 |
| $K_2$(8h) | 0.827447 | 0.343749 | 0.5 | 8 | 0.5877 | $K_2$(8h) | 0.797334 | 0.390163 | 0.5 | 8 | 0.6788 |
| $Fe_1$(4d) | 0 | 0.5 | 0.25 | 4 | 0.0962 | $Fe_1$(4d) | 0 | 0.5 | 0.25 | 4 | 0.0000 |
| $Fe_2$(16i) | 0.284227 | 0.424165 | 0.250213 | 16 | 0.9905 | $Fe_2$(16i) | 0.294942 | 0.402392 | 0.242739 | 16 | 1.0013 |
| $Se_1$(4e) | 0 | 0 | 0.139425 | 4 | 0.9512 | $Se_1$(4e) | 0 | 0 | 0.144973 | 4 | 1.0179 |
| $Se_2$(16i) | 0.092828 | 0.299553 | 0.360999 | 16 | 1.0020 | $Se_2$(16i) | 0.104804 | 0.304507 | 0.354888 | 16 | 1.0046 |
| Rwp = 7.59% ; Rp = 3.98% | | | | | | Space group: I4/m | | | | |

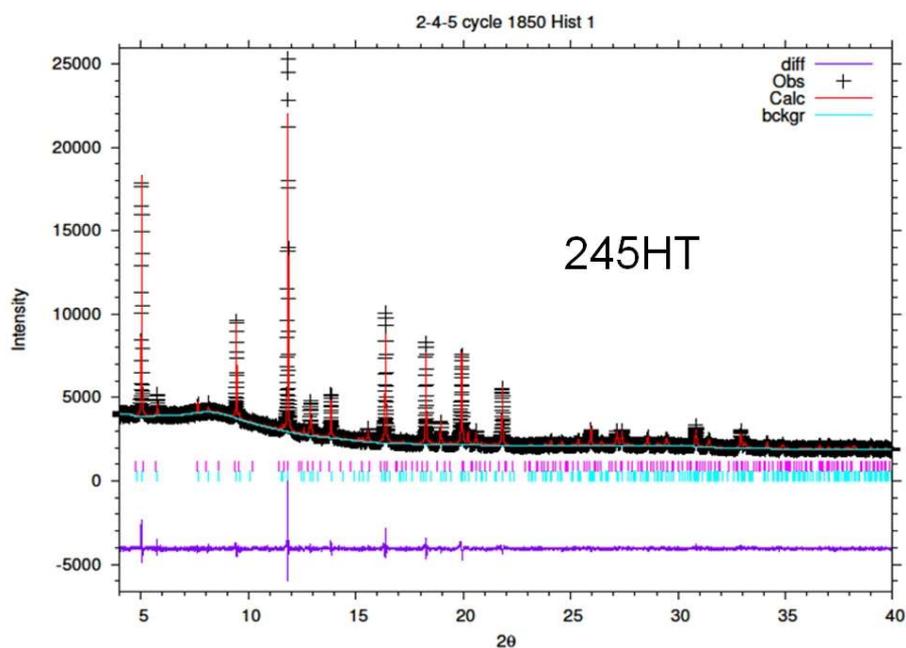

Figure S1. The diffraction pattern of the experimented (red) and simulated (black) results.

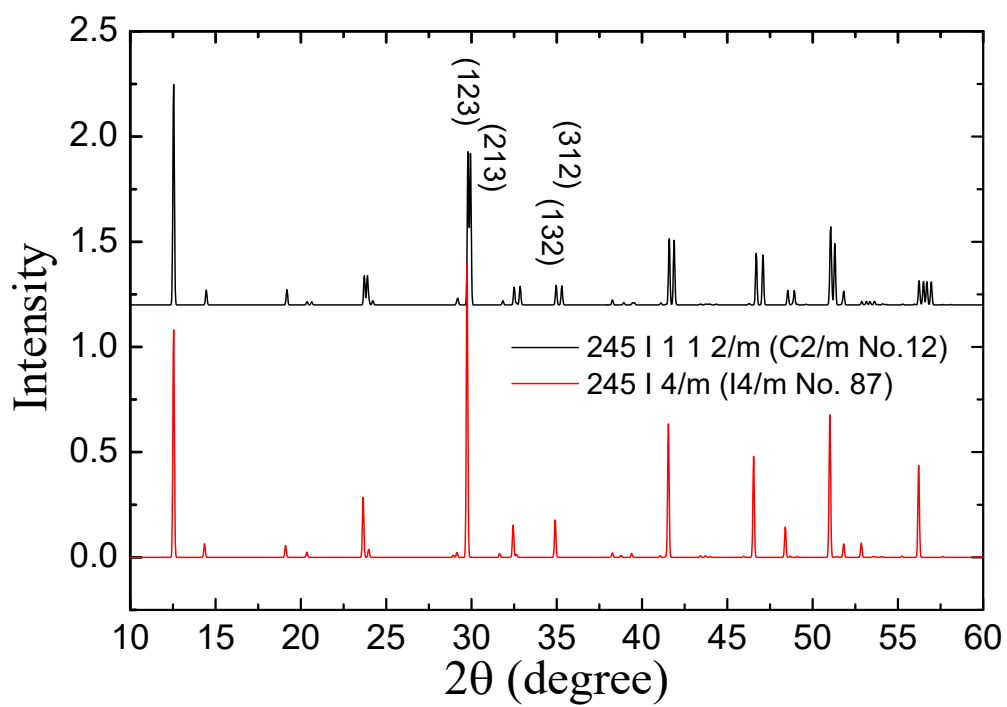